# A 6G White Paper on Connectivity for Remote Areas

April 2020 (*draft release in arXiv.org*)

**Chair**: Harri Saarnisaari, *University of Oulu, FI*, harri.saarnisaari@oulu.fi

**Vice-chair**: Sudhir Dixit, *Basic Internet Foundation, NO, and chair of the CTU in IEEE Future Networks, US*

**Editors**: Mohamed-Slim Alouini, *KAUST, SA*; Abdelaali Chaoub, *National Institute of Posts and Telecommunications (INPT), MA*; Marco Giordani, *University of Padova, IT*; Adrian Kliks, Poznan University of Technology, *PL*; Marja Matinmikko-Blue, *University of Oulu, FI*; Nan Zhang, *ZTE, CN*;

**Other Contributors**: Anuj Agrawal, *IIT Indore, IN*; Mats Andersson, *Forsway Scandinavia AB, SE*; Vimal Bhatia, *IIT Indore, IN*; Wei Cao, *ZTE, CN*; Yunfei Chen, *University of Warwick, UK*; Wei Feng, *Tsinghua University, CN*; Marjo Heikkilä, *Centria University of Applied Sciences in Ylivieska, FI;* Josep M. Jornet, *Northeastern University, US*; Luciano Mendes, *National Institute of Telecommunications, BR*; Heikki Karvonen, *University of Oulu, FI*; Brejesh Lall, *Indian Institute of Technology Delhi, IN*; Matti Latva-aho, *University of Oulu, FI*; Xiangling Li, *Tsinghua University, CN* ; Kalle Lähetkangas, *University of Oulu, FI*; Moshe T. Masonta, *CSIR, ZA*; Alok Pandey, *Indian Institute of Technology (IIT), Delhi, IN*; Pekka Pirinen, *University of Oulu, FI;* Khaled Rabie, *Manchester Metropolitan University, UK;* Tlou M. Ramoroka, *University of Limpopo, ZA*; Hanna Saarela, *University of Oulu, FI*; Amit Singhal, *Bennett University, IN*; Kaibo Tian, *ZTE, CN*; Jun Wang, *Huawei, CN*; Chenchen Zhang, *ZTE, CN*;  Yang Zhen, *ZTE, CN*; Haibo Zhou, *Nanjing University, CN*

**Contents**

# Executive Summary

In many places all over the world rural and remote areas lack proper connectivity that has led to increasing digital divide. These areas might have low population density, low incomes, difficult terrain and non-existing infrastructure, like power grid, making them less attractive places to invest and operate connectivity networks.

6G could be the first mobile radio generation truly aiming to close the digital divide. However, in order to do so, special requirements and challenges have to be considered since the beginning of the design process. The aim of this white paper is to discuss requirements and challenges and point out related, identified research topics that have to be solved in 6G.

This white paper first provides a generic discussion, shows some facts and discusses targets set in international bodies related to rural and remote connectivity and digital divide. Then the paper digs into technical details, i.e., into a solutions space. First, a background overview is provided, followed by a closer elaboration of individual elements, i.e., terrestrial backhaul networks, terrestrial backhaul solutions, non-terrestrial solutions, need for local operations, and frequency spectrum issues. Each technical section ends with a discussion and then highlights identified 6G challenges and research ideas as a list.

The following list provides a high level overview of observed challenges that have to be addressed in 6G remote area research.

| High level view of observed challenges |
| --- |
| Digital divide is increasing, and it is most acute in rural and remote areas. |
| The solution must be affordable and provide sufficient data rate and availability. Furthermore, it should be easy to use and adaptable to different cultures. |
| 6G could be the first mobile connectivity generation that aims for closing the digital divide. To do so, it needs to concentrate on requirements and challenges in rural and remote areas from the beginning of the design cycle. |
| Affordable and sufficient service (data rate and availability) solutions do not call just for technical solutions but also for novel regulation and cooperation between various stakeholders notwithstanding financing challenges. |
| Technically, it uses mobile cellular solutions in places where people live and work (digital oases as we call them) and various backhaul solutions including large cells, relay technology and satellite technology. All solutions should target for affordability and sufficient service, which might differ from targets set for new high data rate solutions for urban, high population areas. |

***About 6G white papers (2020 edition)***

This white paper is one in a series of 6G white papers in 2020. The list of papers is as follows (including cross referencing index at the beginning). *WP1*: 6G Drivers and UN SDGs; *WP2*: Validation and Trials for Verticals; *WP3*: Machine Learning in Wireless Communications; *WP4*: Networking; *WP5*: Broadband Connectivity for 6G; *WP6*: RF & Spectrum; *WP7*: Connectivity for Remote Areas; *WP8*: Business of 6G, *WP9*: Edge Intelligence; *WP10*: Trust, Security and Privacy; *WP11*: Critical and Massive MTC towards 6G; *WP12*: Localization and Sensing.

Draft white papers were published in arXiv.org at the end of April 2020. Final versions will be published in a University of Oulu publication series at the end on June 2020. These white papers are collective efforts of interested contributors from all over the world. They were initiated by 6G Flagship[1] program lead by the University of Oulu in cooperation with 6G Wireless Summit[2].

---

[1] https://www.oulu.fi/6gflagship/
[2] https://www.6gsummit.com/

# Introduction

Today, many rural and remote areas within developed areas around the globe lack reliable, high-quality internet connectivity. Internet access and especially broadband connectivity is even more challenging in developing countries. Indeed, roughly half of world population remains without internet connectivity causing a sharp digital divide. Deficiencies in rural connectivity restrict the use of internet services and adoption of new technologies severely affecting well-being and the economic development in rural and remote areas.

Mobile radio technology generations have increased data rate (bits per second, bps) in each generation as well as total system throughput. Despite the massive potential, rural and remote areas remain largely unserved or poorly served. One of the essential reasons is low expected revenue calculated as average revenue per user (ARPU) which reduces companies' willingness to invest in these areas [1].

Ending the digital inequality calls for changes in investment support strategies and policies as well as in regulation and technology development to create affordable solutions capable of delivering quality of service (QoS) to rural areas comparable with urban areas.

6G has major potential for becoming the first generation to solve global connectivity challenges and dismounting urban-rural injustices as a response to the United Nations' (UNs') sustainable development goals (SDGs) [2]. A reliable mobile network, which has been designed to provide coverage in remote and rural areas, can bring billions of unconnected or poorly connected inhabitants, professionals and entrepreneurs to the information era. High-quality broadband access to applications and services together with connectivity for humans and machines can revolutionize business processes and value chains in rural and remote areas bringing new opportunities for both people and businesses. Towards 2030, 6G connectivity can serve the needs of individuals, households, businesses, governance and machines.

This white paper addresses visions of rural and remote area connectivity to be realised with 6G solutions, with a focus on wireless radio technologies. The ultimate goal is to narrow the digital divide as much as possible with mobile technologies.

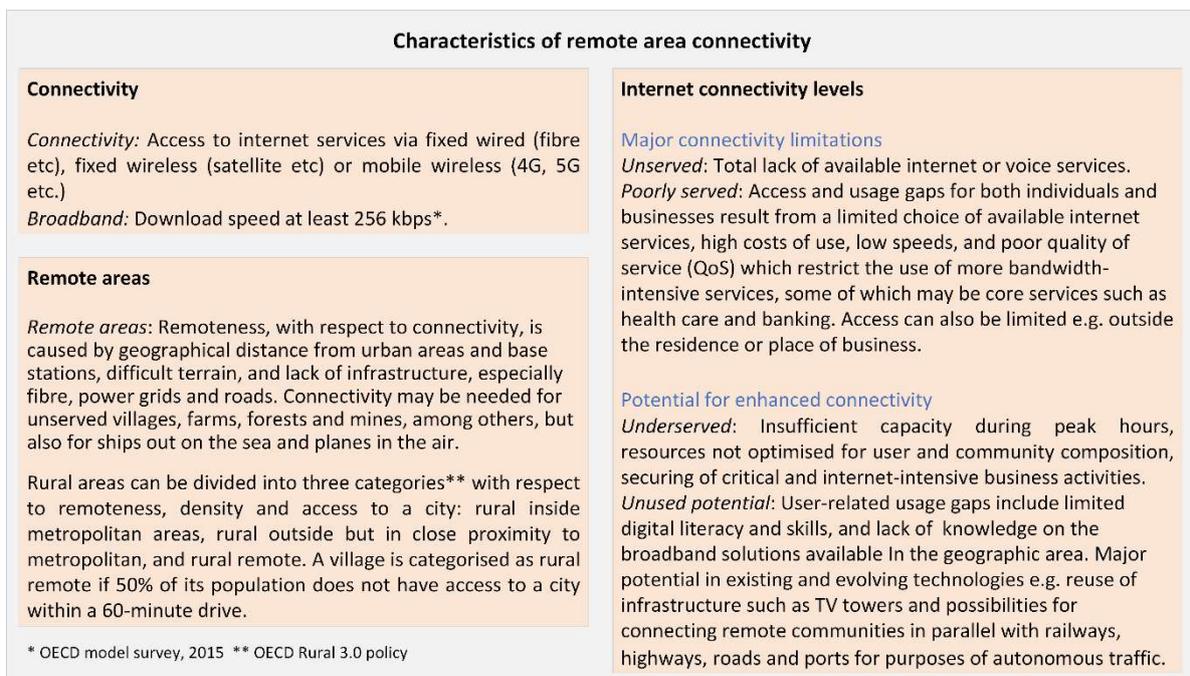

*Figure 1: Characterization of remote areas and a classification of connectivity problems.*

**Key Facts**

In 2018, 55% of global population lived in urban areas [3]. The same year, 67% (5 billion) of world population of 7.6 billion had mobile broadband subscription whereas only 3.9 billion had internet subscription [4] leaving 3.7 billion unconnected. The leading broadband use cases globally are video, fast access to websites (most of them content-heavy) and downloading of large software files (games, updates etc.) [5]. In most cities throughout the world, throughput is 100 – 1000 Mbps with fiber access, or 50 – 500 Mbps with 4G/5G access. However, in rural and remote areas with low population density or low ARPU, the leading use cases are not feasible as they require at least 2 Mbps throughput for a good user quality of experience (QoE). As an example, typical throughput in rural areas in the USA and Europe is approximately 0.5 – 1 Mbps using lines or 3G/4G. In rural India and Africa, on the other hand, typical throughput is as low as 20 – 200 kbps.

In 2018, the UN Broadband Commission for Sustainable Development adopted new targets aiming at connecting the 50% of the world that remains unconnected and to curb increase in the digital divide. One of the targets was to make entry-level broadband service affordable in developing countries, at less than 2% of monthly gross national income per capita [4]. In rural areas in Europe and the USA, reasonable costs per user are estimated to be 30 – 70 USD/month. In rural areas in India, Africa and Brazil, reasonable costs to reach many people are estimated to be 3 – 10 USD/month per user.

**Targets**

The key requirement for rural and remote areas in the next 10 years should be to provide at least 10 Mbps throughput. This will allow everyone to have access like it is today in many urban areas [6]. As a benchmark, the European Commission has set the target (details below) that all households should have at least 100 Mbps connection gradually rising to 1 Gbps. Usually the download speed is of interest in these ranges but in the future, this is expected to change since virtual and augmented reality as well as video services (e.g., in e-health) also increase the uplink speed requirements.

The European Commission adopted a strategy on "Connectivity for a European Gigabit Society" in September 2016, which sets a vision of Europe where "availability and take-up of very high capacity networks enables the widespread use of products, services and applications in the Digital Single Market". Broadband Europe[3] promotes this vision and policy actions to turn Europe into a Gigabit Society by 2025. Three main strategic objectives are included in this vision:

1)  Gigabit connectivity for all the main socio-economic drivers.
2)  Uninterrupted 5G coverage for all urban areas and major terrestrial transport paths.
3)  Access to connectivity offering at least 100 Mbps for all European households.

In addition, 5G connectivity should be available in at least one major city in each member state by 2020 at the latest. The previous broadband objectives for 2020 were to supply every European household with access to at least 30 Mbps connectivity, and to provide half of European households with connectivity rates of 100 Mbps. The European broadband map[4] gives political decision makers as well as private investors the opportunity to monitor the progress made in the deployment of high capacity networks and the quality of broadband services in Europe.

---

[3] Source: https://ec.europa.eu/digital-single-market/en/broadband-europe
[4] https://www.broadband-mapping.eu/

The Broadband Commission for Sustainable Development - a joint effort of the International Telecommunication Union (ITU) and the United Nations Educational, Scientific and Cultural Organization (UNESCO) - has set seven ambitious targets for 2025[5] aimed at connecting everyone, everywhere:

1) By 2025, all countries should have a funded national broadband plan or strategy or include broadband in their universal access and service definition.
2) By 2025, entry-level broadband services should be made affordable in developing countries at less than 2% of monthly gross national income (GNI) per capita.
3) By 2025, broadband-Internet user penetration should reach:
   a) 75% worldwide
   b) 65% in developing countries
   c) 35% in least developed countries
4) By 2025, 60% of youth and adults should have achieved at least a minimum level of proficiency in sustainable digital skills.
5) By 2025, 40% of the world's population should be using digital financial services.
6) By 2025, overcome nonconnectivity of micro-, small- and medium-sized enterprises (MSMEs) by 50%, by sector.
7) By 2025, gender equality should be achieved across all targets.

Target 2 will particularly assist lower income groups in developing and least developed countries to gain connectivity whereas target 6 aims to solve the current challenge of MSMEs that have lower levels of connectivity than large enterprises in the same sectors.

**Previous and Related Work**
From technology standpoint, the survey in [7] provides an excellent overview about technologies discussed in this white paper in addition to the sister white papers. Interested readers should use that for more detailed technical descriptions. 6G technology visions are discussed in [8] and visions in general in the first 6G white paper [9]. Remote area connectivity in the Arctic areas were considered by the Arctic Council in its two task forces on Arctic communication that have reported their outcomes in [10], [11] concentrating on needs, challenges and solutions. Related efforts are ongoing in IEEE Future Networks, especially in the INGR group "Connecting the Unconnected"[6]. Finally, indigenous people have considered connectivity and provided some recommendations in 2019 "Indigenous Connectivity Summit Policy Recommendations"[7], e.g., to use unused spectrum in remote areas for their benefit.

The Organisation for Economic Co-operation and Development (OECD) rural 3.0 policy[8] demonstrates how advances in communications technologies and digital literacy can overcome the challenges of distance through ten key technology areas driving rural change. During the next ten years advanced communications techniques can ensure reliable and high-quality broadband connection for both individuals and businesses enabling the use of internet-based digital services, improving the teleworking experience, and enhancing efficiency of rural businesses. However, access to high-quality broadband alone is not sufficient. Human capital will also need to adapt to the changing technologies while rural areas will need high-quality public services and improved infrastructure - well-maintained airports, roads and ports – to facilitate accessibility and increase attractiveness.

---

[5] 2025 Targets: "Connecting the Other Half"
https://www.broadbandcommission.org/Documents/BD_BB_Commission_2025%20Targets_430817_e.pdf

[6] https://futurenetworks.ieee.org/roadmap
[7] https://www.internetsociety.org/wp-content/uploads/2020/01/2019-ICS-Policy-Recommendation.pdf
[8] https://www.oecd.org/rural/rural-development-conference/documents/Rural-3.0-Policy-Highlights.pdf

OECD's digital economy paper "Bridging the Rural Digital Divide"[9] discusses the challenges of broadband accessibility gaps, good practices for bridging the gaps and approaches for measuring the quality of service offered in each rural or remote area. A focus on download speeds is not sufficient, the paper notes, as individuals residing in rural or remote areas are increasingly becoming producers of content who must also be able to share and create online content while benefitting new enablers such as cloud computing and big data. The paper underlines the importance of tailored targets for broadband availability that take into account the differing capacity requirements depending on the varying number of service users and composition of rural communities – residential, business or anchor institutions. The latter include e.g. hospitals that are carrying out sensitive activities such as telemedicine, and therefore require more intense capacity in terms of bandwidth and reliability.

The OECD's model survey on "ICT Access and Usage by Households and Individuals" as well as the OECD model survey on "ICT Usage by Businesses", both revised in 2015, offer a practical framework for evaluating broadband adoption and barriers to optimal connectivity. Data collected from individuals focuses on a wide array of relevant issues including type of internet and mobile connection; type, frequency and intensity of internet-based activities; and level of satisfaction and perceived obstacles related to access and activities. As an example, reasons for not having access to internet include motivations such as cost of equipment; cost of access; internet services not available or very poor in the area; lack of confidence, knowledge or skills to use the Internet; and privacy or security concerns. Difficulties experienced with mobile connectivity include difficulty in obtaining information on cost of Internet access; unexpected high bills (e.g. due to roaming service); difficulties with mobile network signal (unavailability of broadband or low speed); and difficulties in setting or changing parameters for Internet access (e.g. switching to WiFi, activation of location aware applications or activation of internet access).

Remote and rural areas connectivity is an important study item in different countries around the world. Therefore, there are currently (or just finished) different research projects about the topic. We have identified e.g., EU-Brazil project 5G-RANGE[10], UK project 5G Rural First[11] and EU project ONE5G[12]. In the Nordic region, the Basic Internet Foundation[13] in Norway is a super example of being focused on bridging the digital divide around the globe and addressing the problem of backhaul, both in the developed and developing countries and has developed several solutions for remote areas and villages, and has already deployed them in Norway, Germany and several countries in Africa.

## Playground

Connectivity in remote areas will have a significant impact on people's lifestyle, business opportunities and the society at large. It will reduce the digital divide and improve the state of healthcare, education, transport, agriculture, energy, manufacturing and employment. For example, distance learning and interactive teaching enabled by broadband connection will improve the quality of education. Services such as e-agriculture and weather forecasting may help in improving the productivity in agriculture and minimize the losses. Future may also see popular digital or digitally enabled services in remote areas that we cannot imagine at the moment, but that have their own connectivity requirements. A currently imaginable example is increased usage or virtual and augmented reality in healthcare that may require high speed two-way connectivity.

---

[9] https://www.oecd-ilibrary.org/science-and-technology/bridging-the-rural-digital-divide_852bd3b9-en
[10] http://5g-range.eu/
[11] https://www.5gruralfirst.org/
[12] https://one5g.eu/
[13] https://basicinternet.org/

People often live and work in concentrations what we call *digital oases* in this white paper. These oases could be served using regular cellular technology, e.g., 4G, 5G, and also by WiFi. However, these oases may be away from main roads, outside electric grids or without communication fibres. People would need robust, easy-to-use and secure devices to connect in their oases and elsewhere. Maintenance of the oasis equipment should be simple. People may be poor such that the expected ARPU would be very small. Therefore, new more efficient operation models may be needed to run these connections. Furthermore, user devices should be low in cost. Edge computing and slicing concepts may offer tools to use possible expensive and maybe limited backhaul in an efficient way, e.g., by caching and allowing traffic prioritization based on local needs. These may also be valuable tools to handle possible high variations in data rate requirements, e.g., between working and off-work hours.

Oases must be connected to global Internet using backhauls. Sea and land cables/fibres, microwave links and satellites and other non-terrestrial network (NTN) elements are current solutions, though not viable everywhere and/or to everyone. Global and economically feasible solutions are desperately needed. New emerging low Earth orbit (LEO) satellite systems will be an interesting global opportunity to be followed with great interest in addition to current and new geostationary orbit (GEO) and medium Earth orbit (MEO) as well as highly elliptical orbit (HEO) satellite systems that offer local (though sometime wide area) coverage. Transportation passages connecting oases to other hubs also need connectivity since continuous access is reality in many passages and should also be so in passages of remote areas. It should be kept in mind that polar regions are in different situation since GEO satellites are not covering them such that alternatives are really needed.

People tend to move and work around oases sometimes in difficult environments such as mountain valleys etc. and coverage is needed in such situations as well. Direct user device access by satellite systems or large cells are potential solutions, but currently the usage cost and quite low data rates are the main limiting factors along with the lack of coverage. Ships and airplanes that operate even in more remote environment need connectivity both for operation, maintenance and reporting as well as for staff and passengers.

Frequency regulation in remote places (often) follows nationwide rules though more flexibility could easily be allowed since many frequencies are actually unused in remote locations. Allowed transmit power can be increased to extend range in sparsely populated areas with negligible health effects but delivering significant cost savings. It seems that there is a need for two sets of regulations, one for the urban areas and the other for remote and rural areas. If nationwide operators are not willing to operate in oases, maybe new local or micro operators are needed. Since it would be wise to use the same device everywhere, national roaming or similar solutions should be allowed. Also billing aspects are of interest herein.

We should not forget special needs such as safety and rescue missions that are naturally of interest and would benefit from improved connectivity in remote areas assuming that new systems could fulfil authorities' security, availability and robustness requirements. Furthermore, (environmental) sensing and production related machine-to-machine – Internet of things (IoT) – type communications is an emerging, and often needed, trend. Finally, low power consumption and smart usage of materials and resources as well as sufficient life span and recyclability are indeed factors that have to be considered during development processes.

Communication systems are not standalone solutions but are part of wide technology spectrum in oases. This may affect availability of communication resources. For example, power generators for base stations (BSs) could be a part of the local power grid and availability of power for communication may vary such that its transmission capabilities are changing. These power grids are also needed to change connectivity and other devices.

It is important to understand local culture and to ensur that the community is involved and trusts those offering connectivity services. This is particularly true in the developing world. Solutions to security and authentication should be aligned with the capacity to absorb technologies by people living in remote areas, who some time tend to be local tribes. Thus, along with the technology key performance indicators (KPIs) such as bandwidth, latency, jitter, security, resilience, among others, new KPIs reflecting the rise in economic growth, education, health, gender equality, digital literacy, happiness index, and others in the unserved/underserved remote and rural regions should be taken into consideration for the 6G network design.

Higher data rates are always on demand, but those (even in 6G) may not be well suited for problems described herein. Therefore, a question is, should 6G provide solutions?  We believe that it should. Consequently, in the remaining of this white paper we try to envision solutions to these problems and raise key challenges and research questions around the topic. At the end, economic factors as well as nations' and companies' ambitions will show what could appear in 2030s, in the 6G era.

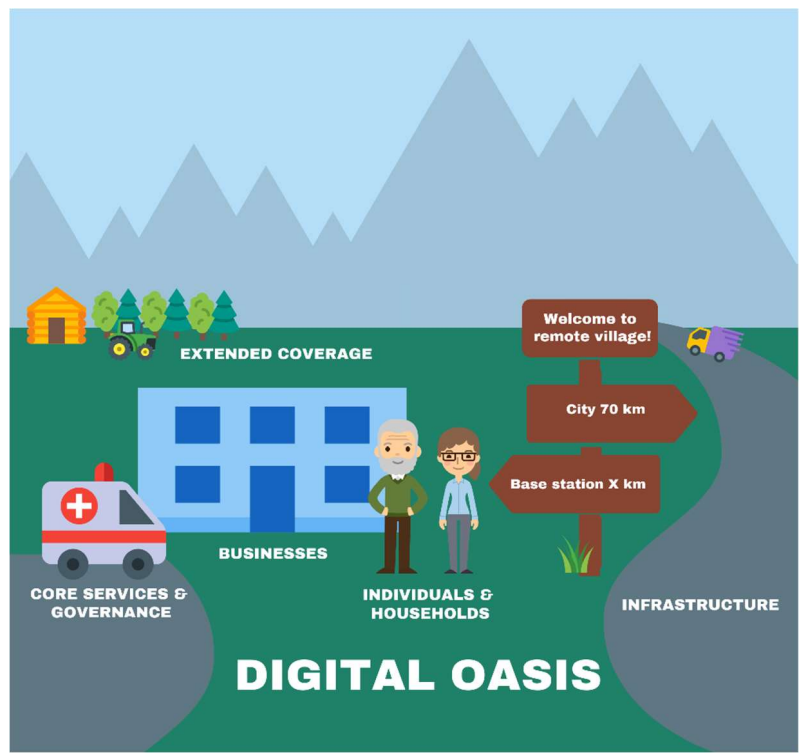

*Figure 2: An illustration of the playground in this remote area 6G white paper.*

| **6G Challenges and Research Ideas** |
|---|
| • Affordable, low cost solutions/technology. |
| • Not just coverage, but also sufficient 24/7 two-way data rate (QoS). |
| • Solutions must be environmentally friendly and allowing circular economy. |
| • Cultural aspects may affect how technology and possibilities it provides are accepted and used. |
| • Specific remote area requirements including e-farming and related IoT systems. |
| • Adaptive system for power supplies to autonomous base station within local micro power grid. |
| • Key drivers to support operation in low ARPU areas. |

# Wireless Terrestrial Mobile Radio Solutions

In this chapter, we focus on wireless terrestrial mobile radio solutions to deliver wireless connectivity or Internet access in rural and hard to reach areas.

Mobile terminals seem to be the last-mile solution also in the near future (also in 2030s) to provide all kind of connections in areas that lack fixed connections. Consequently, future mobile solutions should be affordable. The cellular infrastructure and mobile terminal technology should be low cost, and also usage (along with backhaul costs) should be cost effective (or subsidised somehow) such that people in low income areas could use them. For achieving this target, along with constant improvement for a higher data rate in 6G solution, a separate lower cost solution might be needed with unified device types. Meanwhile, these oases serving mobile solutions should opportunistically use all available backhaul services.

Some oases (and relay stations) are outside power grids. Therefore, they must rely on refuelled generators or renewable energy such as solar, wind and water. Hence, low power consumption is a one development driver and careful system optimization based on power availability and usage may be needed. Furthermore, a power plant in an oasis may be a part of a local power grid used also for household needs. This calls for a new type of (joint) resource optimization both in communication systems and power grids.

One of the immediate approaches to deliver wireless access to hardly-accessible areas is to create extremely-large coverage cells, also known as mega-cells. Deployment of new BSs for mega-cell creations for wireless access delivery is often associated with some investment costs, which may be reduced if existing infrastructure such as high TV towers and buildings in hills/mountains as well as new infrastructure such as tethered balloons could be used.  These high-spot transceivers can be used both for end user access as well as to provide backhaul connection to oases. These high-spot devices could also serve as relay stations in a microwave link chain.  The future 6G system should support this type of multipurpose use and large coverage that may need some investigations. In addition, suitable antenna technology may require some research attention in 6G frequencies.

Another pillar in the domain of terrestrial networks application for wireless access delivery in difficult to reach locations is the utilization of so-called moving platforms, as well as the incorporation of the data caching concept.  It is envisaged that in the future more and more services are non-human originated. Some of those require real-time communications, but many IoT type services have very relaxed delivery time requirements. In such cases, data can be collected using moving platforms. The class of moving platforms include not only nomadic and movable base stations mounted on dedicated platforms, often (semi-) autonomous, but also vehicles, human users, unmanned aerial vehicles (UAVs), and other non-terrestrial platforms. Those can be used also to distribute information and update software etc. at the local nodes enabling e.g., caching based services locally. Connectivity to moving platforms can be rather random in nature and the basic operation of nodes should not depend on them.

Slicing and caching can together also play a major role in rural and remote connectivity as local content creation and provision is envisioned. Moreover, the cacheable data in rural and remote areas may be higher as compared to the urban regions. Therefore, local community will need to be empowered to create their own content and store it in local server. Network slicing will enable resource-efficient realization of content-based connectivity and creation of differentiated service environment including economic/free and premium/paid. Slicing and caching also help to keep local traffic local and offer other means to avoid using possibly limited and costly backhaul connection.

In many scenarios, allocated frequencies are used in urban areas and along main road. Since frequencies may be allocated nationwide, some allocated frequencies remain unused in remote areas. Relaxing allocation rules and allowing local frequency usage, while keeping possible interference issues in mind, might ease improving connectivity in remote places.

Sometimes connectivity is temporarily needed in areas where it does not exist or is not available anymore, e.g., due to a disaster. In this case ad hoc networks are needed and often the need is by "blue light" authorities such as police officers, firefighters and first responders. Deployable networks are such ad hoc solutions. They can be used to form independent local networks, but also connected to internet many opportunistic ways. However, while developing 6G solutions for these, it is important to understand that public protection and disaster relief (PPDR) authorities may have specific requirements usually not needed in other uses. See [12] for further discussions.

Sometimes people need to be educated and accept new technology. An example of how to do this is gaming and quiz apps developed for a particular rural region, where the villagers self-learn the lessons on how to utilize the technology and devices through interactive audio-visual games and quizzes based on multiple choice questions.

| 6G Challenges and Research Ideas |
|---|
| • Technology and standards for large cell solutions.<br>    o Possible coexistence of 6G and TV signals in a TV tower.<br>• Power usage adaptation based on available power from a local power grid taking other local power consumption into account.<br>• Tailoring caching and slicing for remote area use cases, both user and IoT.<br>• Mobile platforms, e.g., in collecting sensor data and supporting connectivity.<br>• More relaxing and flexible frequency regulations in remote places along with interference management issues.<br>• Deployable 6G networks, e.g., for emergency and authority use. |

## Wireless Backhaul Solutions within 6G

Network deployment in rural areas (i.e., the most under-connected areas) is complicated by the varying degree of terrain that may be encountered when installing cables or fibres between cellular stations. Recently, the research community has also started investigating the feasibility of integrated access and backhaul (IAB), in which only a fraction of BSs connects to traditional fibre-like infrastructures, while the others wirelessly relay the backhaul traffic, possibly through multiple hops [13].

Although Long Term Evolution (LTE) and LTE-Advanced already support BSs with wireless backhaul, future IAB developments foresee a more advanced and flexible solution, with multi-hop communications, dynamic resource multiplexing, and a plug-and-play design for low-complexity deployments. Additionally, a wireless backhaul makes it possible for non-terrestrial platforms, including satellites, to serve backhauling requests from on-the-ground terminals, thereby saving terrestrial resources for access operations. The importance of the IAB framework as a cost-effective alternative to the wired backhaul has been recognized by the 3GPP, within constant optimization over Releases 15 to 17 on architectures, radio protocols, and physical layer aspects for sharing radio resources between access and backhaul links [14]. Specifications may continue with enhancements and other scenarios in future, i.e., as part of beyond-5G standardization efforts.

IAB presents lower deployment costs and complexity compared to traditional networks with fibre backhaul and facilitates the site installation even where fibre may not be available (e.g., rural areas have a varying degree of terrain, making executing a cable or fibre buildout between cellular towers even more difficult and expensive). The potential of the IAB paradigm is particularly evident when wireless backhaul connections are realized at mmWave frequencies, thus exploiting a much larger bandwidth than in legacy sub-6 GHz systems. Moreover, IAB in mmWave offers the possibility to multiplex the access and backhaul data within the same frequency band, thereby removing the need for additional hardware and/or spectrum license costs.

Along these lines, the massive data rate requirements of the new 6G access technologies may require further growth of the backhaul capacity. In this direction, utilizing the THz band becomes quite important. Traditionally, the lack of high-power compact transmitters and high-sensitivity receivers, combined with the high propagation loss resulting from the much smaller wavelength at THz frequencies, limited the communication distance in the THz band. However, recently, major advances in THz device technologies [15] and tailored physical-layer techniques are making long-range THz communications a reality, not only for on-the-ground backhauls, but even for satellite communications [16]. While there are many challenges accompanying this spectrum, including the impact of atmospheric effects and the need for passive/active service co-existence, we expect THz IAB to be part of 6G future studies.

At the same time, it will be extremely important to design ad hoc scheduling procedures, to efficiently split the resources between the access and the backhaul and provide interference management, so as to avoid worse network performance in congested networks. Another important element to be considered in an IAB architecture will be the establishment and management of the network topology, including network formation, route selection, and resource allocation. In particular, the IAB topology should be dynamically adapted when a backhaul link is degraded or lost (to maintain service continuity) or when congestion arises [17]. Also, retransmission and packet reordering in case of multi-hop IAB connection retransmissions may introduce additional delays and have a negative impact on end-to-end performance.

Finally, it should be mentioned that the IAB architecture may require a certain number of optical fibre infrastructures to be already deployed. In this regard, fibre backhaul capacity can be increased if the existing wavelength division multiplexing based network is gradually migrated to elastic optical network by technology upgradation at nodes, without the need of expensive deployment of new optical fibres. Moreover, the rich power distribution network infrastructure should be leveraged to improve connectivity in the remote regions without significant investment by using either the overhead fibre cable network of power-grid companies or using the cost-effective power line communication technology, both narrowband and broadband. Likewise, road/highway lights could be opportunistically used whenever feasible.

Although this section was about IAB, also other advances in micro-wave radio link technology are welcome.

| 6G Challenges and Research Ideas |
|---|
| • Advanced backhaul connections, e.g., <br>      o Via mmWave and THz frequencies including novel antenna solutions. <br>      o Visible line and power line backhaul solutions. <br>      o Satellites, mega-cells. <br> • Flexible IAB application: <br>      o Scheduling procedures that dynamically split the resources between the access and the backhaul. <br>      o Efficient routing/path selection strategies for the determination of the optimal backhaul path which are robust to network topology changes and end terminals' mobility. <br> • Smart (re)usage of existing infrastructure. |

## Non-Terrestrial Network Solutions

Satellites and high frequency (HF) solutions are currently applied in very remote areas (sea, air, desert), and are likely to be the case in their future versions. Non-terrestrial network (NTN) systems include satellite systems (SATCOM) at different orbits, high altitude platforms (HAPS), UAVs as well as balloons. NTN solutions may also serve other remote places and oases by providing direct user access or backhaul connection or complementing terrestrial services. At the end, we should look for hybrid or integrated terrestrial and NTN solutions.

The benefits of NTN include [18]:

- *Communication resilience:* Non-terrestrial platforms enable wide connectivity coverage and guarantee seamless service continuity, e.g., in rural areas or when terrestrial infrastructures are not available as on oceans.
- *Energy-efficient connectivity:* Non-terrestrial nodes can be deployed on-demand implementing smart duty cycle control mechanisms, thereby reducing operational and management costs of fixed infrastructures.
- *Resource optimization on parallel backhaul links:* Non-terrestrial platforms offer an additional and robust channel for backhauling operations, thereby saving terrestrial resources for access traffic requests. This also guarantees that the on-the-ground terminals can find an alternate route to preserve the connection if terrestrial links are unavailable.
- *QoS enhancement through edge computing:* Air/spaceborne stations, including satellites, can host mobile edge cloud functionalities to support communication, computing, and storage operations for on-the-ground users to execute their cloud services
- *Communication on the move.* Satellites provide high-speed connectivity to individual in-motion terminals that cannot benefit from terrestrial coverage, such as planes or vessels.

A disadvantage of existing satellite systems is the high cost of both user terminal (e.g., within BSs) and satellite connection. Future solutions should address these and provide economically affordable solutions.

Efforts are ongoing in satellite sector for improved capacity, both on more traditional GEO/MEO orbits, and on emerging concepts in LEO /HEO orbits that have also polar coverage. Some balloon initiatives exist too; see [7] for further details. Technology advancements that have made this possible include Gallium Nitride (GaN) technology, emerging mass production of satellites, antenna arrays for tracking multiple mobile satellites and intersatellite (optical) links (ISL) for enhanced spaceborne routing meaning that ground gateways are not always needed. Inter-satellite system connectivity may further improve the situation as well as inclusion of terrestrial systems into this system-of-systems. Free-space optical technology as well as other new radio frequencies, e.g., mmWave, may be used for connecting satellites to their ground gateways with even higher data rates.

Integration of NTN with mobile terrestrial systems has been started as a 3GPP study item with the expectation to reduce the cost with common chips in addition to other benefits. After identification of the possible challenges, 3GPP has started to solve these problems [19]. Since especially satellite system design lifecycle is rather long, it is anticipated that these 5G (Release 17 and 18) based solutions will be in use in early 2030s. 6G should include the NTN component in its design from the beginning so as to avoid spending a lot of effort into this afterward. Integration is considered also by satellite actors, e.g., in [20].

Among all innovations, the availability of multi-layered networks [21], [22], i.e., the orchestration among different aerial platforms operating at different altitudes, currently represents one of the most promising technological options for non-terrestrial systems and makes it possible to obtain better spatial and temporal coverage. The service model envisaged in this regard comprises of two configurations. In the first arrangement, a single non-terrestrial platform operates in a "tower-in-the-air" configuration whereby it relays data obtained from the ground station (uplink) to various service delivery platforms in the downlink. In the second configuration, a swarm/cascade of aerial stations is used as both relay nodes and service delivery devices for the local users. Both LEO and MEO satellites, as well as HAPS, can be used together if the area to be covered is significantly large by combining multiple radio technologies into a single solution that is more robust and efficient than any individual approach. Despite current standardization efforts towards the development of NTNs [19], constant optimization for proper protocol design still calls for long-term research to overcome the larger propagation, coverage, high mobility and enhance the service quality and service continuity. Also, from implementation perspective, a constellation to maintain ubiquitous service continuity, which may in turn be very difficult to install.

Moreover, for ensuring the smooth commercial usage of future integrated network, closely cooperation among different operators with infrastructure sharing may also be needed. For examples, the core network constructing by the terrestrial operator can be shared with the satellite based access for user equipment (UE) located at different regions. A unified set of services can be provided with adjustment per access based on radio access network (RAN) capability. Meanwhile, such mechanism with specific gateway is feasible for the integration between TN and NTN with either unified or different access techniques.

From the above discussion, it appears that, even though non-terrestrial networks are emerging as a key component of the future 6G telecommunication landscape, there are still various optimization needed to be done in further research.

| 6G Challenges and Research Ideas |
|---|
| <ul><li>From standardization perspective, to enable the integration of NTN and TN (e.g., at beginning of 6G design) including:<ul><li>Enhancement on physical, access and network layers address the challenges imposed by the different NTN channel profile, e.g.,<ul><li>deal with the very large end-to-end propagation delay;</li><li>satellite friendly signals (e.g., low back-off);</li><li>flexible design to achieve the concept of "routing" for the internet of space including local connectivity via NTN;</li><li>trade-off between the coverage vs. data rate.</li></ul></li><li>Enable MEC via NTN node.</li><li>Resource utilization optimization/sharing in hybrid systems even if non-6G (non-3GPP) NTNs are used.</li></ul></li><li>Optimization of multi-layered networks.</li><li>From implementation perspective, to enable the NTN with<ul><li>efficient power amplifiers for satellites to boost their efficiency and improve link budget;</li><li>efficient (programmable) satellite platforms and stability to overcome the difficult on maintenance and reduce the cost within the life of the satellite.</li></ul></li><li>Low cost NTN solutions/technology and affordable usage.</li></ul> |

## Micro Operators

So far leading operators have failed to provide sufficient connectivity for many remote places all over the world. Recently, governments have either supported in this or "directed" operators to co-operate while coverage is created. Despite of these developments new operational models may be needed in order to further increase the penetration of Internet coverage. The establishment of local networks without a direct involvement of national operators has become increasingly important to serve areas where operators do not see a business case. We call them local private networks or micro operators in this white paper. It means that a village or property owner could run its own network in order to provide connectivity in the area. Naturally, this has to be done according to regulations and rules, and heavily depend on the local availability of spectrum. In remote areas, infrastructure building may be supported by other bodies such as private companies, governments or international organizations. Furthermore, co-operation with leading operators should be guaranteed. Micro operators who engage with the community and come from the same stock have much greater chance of success as they take into consideration their local needs and can even leverage their infrastructure to reduce the capital and operations cost to become profitable.

The potential of micro operator concept has been partially illustrated by successfully deploying it in Peru, to improve the rural area connectivity[14]. In that case the micro operator, who deploy network infrastructure and operates it in rural areas, is called the rural mobile infrastructure operator (RMIO), whose infrastructure is interconnected to mobile network operator's (MNO's) network core and revenues are shared.

Many nations do execute rather rigid frequency regulation where frequencies are often given through auctions nation-wide to one or a few operators, indicating that micro operation by non-MNOs is indeed rather impossible due to the lack of spectrum available for them. Once lighter regulatory requirements for such local micro operators and a supportive government regime is shown beneficial and if more flexible frequency use would be allowed, unused spectrum in remote areas could be utilized. However, if this is not an internationally agreed process, the market would be small, and devices will be expensive unless new flexible low-cost radio platforms could be available. Flexible frequency use means that interference control (in spectrum) must be automated rather than executed manually. For a sustainable micro operator ecosystem, issues such as maintenance, risk, resilience, interference, scalability and power need to be considered, and these challenges can be most effectively and efficiently handled through community training and involvement.

Users in oases should also be allowed to join larger networks and vice versa. Therefore, national roaming or alike solution should be ensured. Furthermore, sensible billing procedures are needed for micro operators. In addition, means to control traffic share has to be developed. Local networks might have limited capacity such that a (large) number of visitors may totally block the system. Therefore, local needs have to be guaranteed. This could be managed by having a network slice for locals and visitors and dynamically adjusting the bandwidth based on load but by guaranteeing locals their share if they need it. Finally, local communities should be able to manage what are the most important applications for them, e.g., do they favour e-learning and e-health services before video browsing such that possible limited backhaul could be used the most efficient way. All this in addition to keeping local traffic local and utilizing edge computing for caching (also to reduce backhaul usage) that should be available after 5G.

---

[14] 5G-RANGE project, "Deliverable D7.1 -Exploitation, communication, dissemination and standardization – Part I," January 2019. http://5g-range.eu/

To ensure high QoS for the services being provided, the operators and the local community should be made skilled enough to maintain and repair the equipment used in the deployed technology. They should view learning digital technologies as pathways to better employment and well-being for the community.

| 6G Challenges and Research Ideas |
| --- |
| <ul><li>The whole local micro operator ecosystem must be considered. It includes various aspects such as<ul><li>available frequencies: new flexible frequency regulation utilizing unused spectrum in remote areas and allowing local operation;</li><li>national roaming (or alike solution) to allow nationwide use of the service agreement;</li><li>billing system;</li><li>maintenance and operation;</li><li>financing models;</li><li>dual use with government actors such as SAR, healthcare, education, etc.</li><li>Co-existence and co-operation with leading operators.</li></ul></li></ul> |

## Radio Frequency Related Issues

The radio spectrum is the lifeblood of the telecommunications market and the debate centers mainly on how this resource, despite being limited and naturally finite, can drive both urban and remote connectivity. Academic and industry communities identified two generally accepted solutions to be able to achieve the 6G promise of everywhere connectivity. Either to allocate additional spectrum through sharing/refarming or introducing new frequencies mainly in the higher bands, which also increasingly involves sharing/refarming. This is a joint effort between researchers, standardization bodies, operators and regulators.

Nowadays, radio spectrum is exclusively and statically allocated to different wireless services which leaves very limited spectrum for fixed frequency assignments. This has resulted in a situation that spectrum shortage is a result of the outdated spectrum management policy rather than physical scarcity [23]. More flexible spectrum usage is preferred in future.

Sharing methods have been studied in the past and some have even been applied. These efforts should be continued during 6G research and further developed. Especially, as highlighted in this white paper, techniques that ease operation in remote areas are most welcome. These include local micro operators, cognitive radio networks, coexistence with sub-6GHz bands and multi-tiered spectrum access methods.

Meanwhile, mmWave and THz bands can provide large chunks of spectrum, needed to meet the extremely high throughput demands in 6G. However, these bands are very sensitive to molecular absorption and impose tight constraints in terms of line-of-sight propagation with very limited penetration capabilities. On the positive side, these characteristics makes mm and THz bands perfect for opportunistic spectrum reuse. Future research efforts can focus on harmonious coexistence of users as the communications are short-range and highly directional and thus interference can be contained by avoiding the transmit beams or just by being located a few meters away. To attain this objective, novel beam management and power level coordination schemes need to be designed. Unfortunately, the flip side of using these bands is that the infrastructure investment required is quite significant and the operator must make difficult trade-offs between coverage, capacity and cost.

| 6G Challenges and Research Ideas |
| --- |
| <ul><li>New and more flexible regulatory framework.</li><li>Innovative incentive mechanisms to motivate the incumbents to share their underutilized or unused spectrum especially in attractive frequencies such as the lower and mid bands. In this regard, infrastructure sharing agreements between operators could be inspiring.</li></ul> |

> - o  Adopting new business models and regulatory proposals for spectrum access frameworks that are technically feasible and commercially affordable and support remote area connectivity.

## Summary

The goal of the white paper was to identify 6G research ideas as they relate to connecting the remote areas. These ideas can be used to propose further research and development in 6G to connect everyone on this planet, especially those still under-connected in remote and rural areas. With primary focus on the urban areas, the issues highlighted in this paper have not received the much needed attention. The ideas and challenges were summarized in each section. In summary, there is no shortage of excellent research ideas to be able to serve the remote areas. These research ideas span across multiple stakeholders, from users to operators to regulators to content providers to local entrepreneurs.

The following figure illustrates the technical areas that were discussed in this white paper in the context of connecting the remote and rural areas to achieve the vision of ubiquitous connectivity.

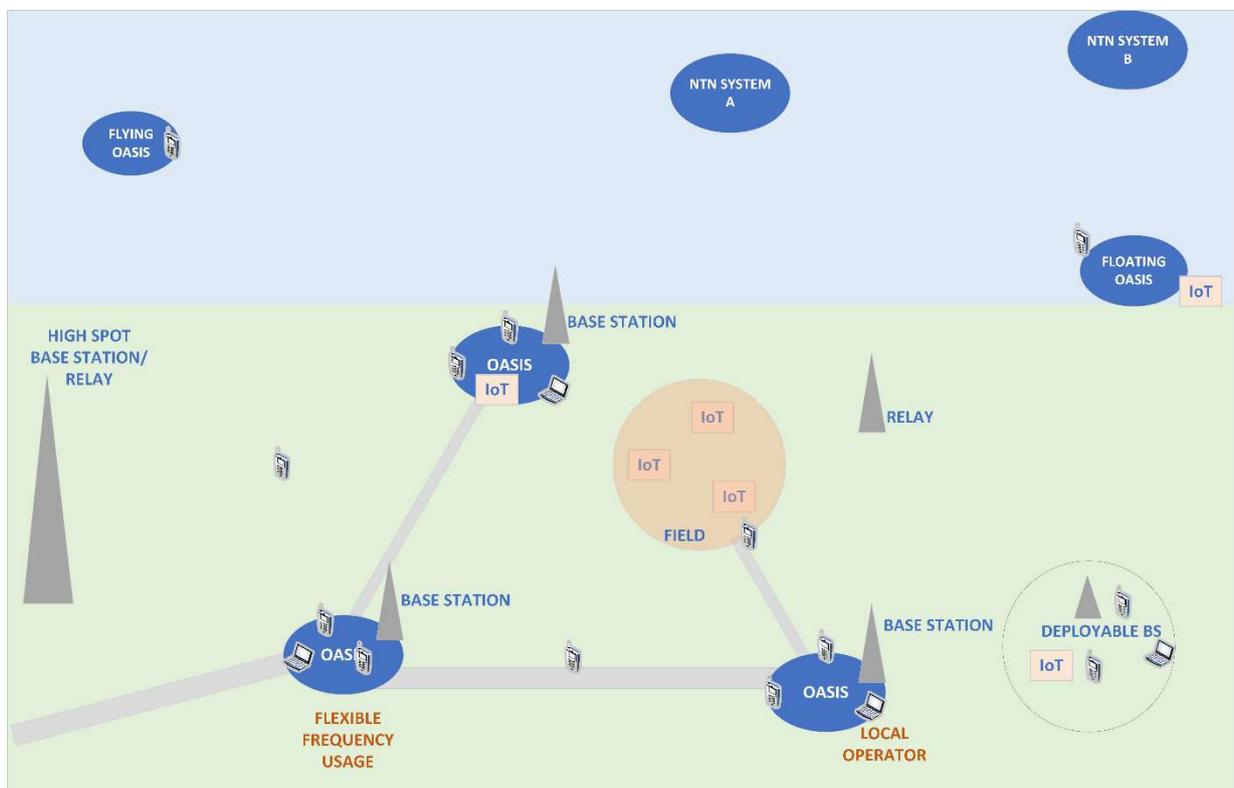

*Figure 3: Illustration of the broad landscape that drive the various elements of the overall technical solution.*